\documentclass{appolb}
\usepackage{epsfig,amssymb,amsmath}

\def\bge{\begin{equation}}
\def\ene{\end{equation}}
\def\bg{\begin{eqnarray}}
\def\en{\end{eqnarray}}

\def\bge{\begin{equation}}
\def\ene{\end{equation}}
\def\bg{\begin{eqnarray}}
\def\en{\end{eqnarray}}

\begin{document}
\title{
\hbox{
QED and Fundamental Symmetries in Positronium Decays
}
}
\author{Steven D. Bass 
\address{
Marian Smoluchowski Institute of Physics, 
Jagiellonian University, \\ 30-348 Krakow, Poland \\
\vspace{3ex}
Kitzb\"uhel Centre for Physics, Kitzb\"uhel, Austria}
}
\maketitle
\begin{abstract}
\noindent
We discuss positronium decays with emphasis on tests of fundamental symmetries and the constraints from 
measurements of
other precision observables involving electrons and photons.
\end{abstract}

\section{Introduction}

Precision studies of positronium decays are a sensitive 
probe of QED and allow new tests of fundamental symmetries 
involving charged leptons.
Prime topics for experimental investigation
include decay rates,
tests of discrete symmetries and
looking for possible rare and invisible decays of 
positronium atoms.
With a new generation of positronium experiments coming 
on-line \cite{Moskal:2016moj,mitps,Crivelli:2010bk}
it is valuable to review 
the present status of theory and
constraints from other processes on the key observables.
This we present here, with main emphasis on physics
experiments~\cite{Moskal:2016moj,Kaminska:2016fsn,Gajos:2016nfg}
that can be performed using the new J-PET detector,
Jagiellonian Positron Emission Tomograph,
being built in Cracow~\cite{Moskal:2014sra}.
The J-PET is a new PET device based on plastic scintillators
designed for total body scanning in medicine
as well as
biological applications~\cite{Moskal:2016a} 
and fundamental physics research~\cite{Moskal:2016moj} 
with detection of positronium and (from its decay products)
Compton rescattered photons in the detector.

The physics of positronium 
(an ``atom'' consisting of an electron and a positron) 
is described by QED 
with small radiative corrections from 
QCD and weak interaction effects in the Standard Model.
Positronium comes in two ground states,
$^1 S_0$ para-positronium, denoted p-Ps, 
where the spins of the electron and positron are antiparallel
and 
$^3 S_1$ ortho-positronium, 
denoted o-Ps, 
where the spins of the electron and positron are parallel.
\hbox{p-Ps} is slightly lighter by 0.84 meV due to 
the interaction between the electron and positron spins 
and also the existence of virtual annihilation processes.
Spin-zero p-Ps has a lifetime of 125 picoseconds 
and spin-one o-Ps has a lifetime of 142 nanoseconds.
Reviews of positronium physics are given in
\cite{Cassidy:2018tgq,Karshenboim:2005iy,Gninenko:2002jn}.

Measurements of positronium decay rates are consistent 
with QED theoretical predictions
although the present experimental uncertainties 
$\sim {\cal O} (10^{-4})$
are very much larger than the theoretical uncertainties
on the QED calculations,
by a factor of 100 for o-Ps and by 10,000 for p-Ps,
calling for increased experimental precision.

Precision observables in positronium decays can be used 
to test discrete symmetries
$C$, $CP$ and $CPT$
with charged leptons.
Charge conjugation invariance here has been tested
up to the level of $10^{-6}$ \cite{X5,X4,X2}.
The symmetries $CP$ \cite{Yamazaki:2009hp,Y1}
and $CPT$ \cite{Y2,X1}
have each been tested up to ${\cal O}(10^{-3})$.
QED final state effects in o-Ps decays
can mimic $CP$, $T$ and $CPT$ violation at
the level of $10^{-9}$ to $10^{-10}$
\cite{Bernreuther:1988tt,Y2}.
Possible invisible decays of positronium are also an interesting topic of investigation.
Mirror matter models of dark matter allow a branching 
ratio for the invisible decay of o-Ps in vacuum 
to mirror particles
up to about $2 \times 10^{-7}$, 
below the present experimental bound of $5.9 \times 10^{-4}$
\cite{Vigo:2018xzc}.

Possible ``new physics scenarios'' 
are strongly constrained by studies of other 
(QED related) 
observables including precision measurements of 
the fine structure constant
\cite{Hanneke:2008tm,Parker:2018sc}
and the electron electric dipole moment (EDM)
\cite{Andreev:2018ayy}.
Here we discuss the status of this physics 
taking into account the 
latest from theory and experimental investigation.
Section 2 describes positronium decays in QED.
In Section 3 we discuss precision measurements of 
$\alpha$ and the electron EDM, and 
their constraints on possible positronium decays.
In Section 4 we explore discrete symmetry tests as
well as rare and exotic decays,
e.g. involving axions and possible invisible decays.
Conclusions are given in Section 5.

\section{Positronium decays in QED}

\subsection{o-Ps decay rate}

Positronium properties such as decay rates and 
energy levels can be calculated using 
the formalism of non-relativistic QED~\cite{Labelle:1992hd}. 
The o-Ps decay rate within QED has been evaluated to two-loop level.
One finds \cite{Adkins:2002fg}
\begin{eqnarray}
\Gamma( \hbox{o-Ps} \to 3 \gamma, 5 \gamma )
&=&
\frac{2 (\pi^2 - 9) \alpha^6 m_e}{9 \pi}
\biggl[ 1 
+ A \frac{\alpha}{\pi} 
+ \frac{\alpha^2}{3} \ln \alpha 
\nonumber \\
& & \ \ \ 
+ B \biggl( \frac{\alpha}{\pi} \biggr)^2
- \frac{3 \alpha^3}{2 \pi} \ln^2 \alpha
+ C \frac{\alpha^3}{\pi} \ln \alpha
+ D \biggl( \frac{\alpha}{\pi} \biggr)^3
+ ...
\biggr].
\nonumber \\
\end{eqnarray}
Here
$A = - 10.286606 (10)$,
$B= 44.87 (26)$ if just the 3 $\gamma$ decay is included
and
$B = 45.06 (26)$ if we also include the $5 \gamma$ decay,
$C = A/3 - 229/30 + 8 \ln 2 = -5.517025 (03)$
and $D$ remains to be calculated.
There is good convergence of the perturbation expansion. 
The terms in Eq.~(1) evaluate as
7.211167,
-0.172303,
-0.000630,
0.001753 (11),
-0.000032,
0.000024,
0.00000009 D, 
-- see Table XVII of \cite{Adkins:2002fg}.
That is, 
the NLO (next-to-leading order)
term proportional to $A$ is 2\% 
of the leading Born term
and the NNLO term proportional to $B$ is just 0.02\%.
(Expressions for $A$ and $C$ in closed analytic form
 are given in \cite{Kotikov}.)
The $5 \gamma$ decay contributes
$ 0.19(1) \bigl( \frac{\alpha}{\pi} \bigr)^2 
\approx 10^{-6}$
through the $B$ parameter 
in Eq.~(1)~\cite{Lepage:1983yy,Adkins:1983tu}.
QED light-by-light contributions 
to the positronium decays \cite{Adkins:2001zz}
contribute
0.350(4) to the $B$ parameter 
for o-Ps in units of 
$\bigl( \frac{\alpha}{\pi} \bigr)^2$
times the respective lowest order rate.

Including both the 3 and 5 photon contributions gives 
the QED decay rate prediction
\begin{equation}
\Gamma = 7.039 979 (11) \times 10^6 s^{-1}
\end{equation} 
where we here neglect the ${\cal O}( \alpha^3 )$ term 
proportional to the unknown constant $D$~\cite{Adkins:2002fg}.

The calculations in \cite{Adkins:2002fg} are pure QED.
There will also be hadronic 
QCD light-by-light and
photon self-energy corrections.
Hadronic light-by-light corrections are 
known to be important in 
the muon $g-2$ puzzle \cite{Jegerlehner:2017gek}.
In positronium decays these corrections 
will enter at 
${\cal O} (\alpha^2 )$ in the decay rate
with extra suppression factor 
$(m_e/\mu)^2 \sim 10^{-6}$ in the $B$ parameter
with $\mu$ a typical hadronic scale,
and are well beyond present experimental reach.

The most accurate measurements of o-Ps decays are consistent
with each other and with the theoretical prediction, Eq.~(2).
Kataoka et al.~\cite{Kataoka:2008hj} found
\begin{equation}
\Gamma = 7.0401 \pm 0.0007 \times 10^6 s^{-1}
\end{equation} 
with the o-Ps produced in SiO$_2$ powder,
whereas Vallery et al.~\cite{Vallery:2003iz} 
found
\begin{equation}
\Gamma = 7.0404 \pm 0.0010 \pm 0.0008 \times 10^6 s^{-1}
\end{equation}
working in vacuum.

These results are consistent with the QED 
theory prediction, Eq.(1) with the caveat 
that the present experimental uncertainties 
on the decay rate 
are about 100 times greater than the theoretical error.
The leading ${\cal O}(\alpha)$ correction in Eq.~(1)
is needed to agree with the data, 
the ${\cal O}(\alpha^2)$ 
terms are of order the same size as
the experimental error and 
the ${\cal O}(\alpha^3)$ terms are
well within the experimental uncertainties.
Five photon decay measurements are consistent 
both with zero and with theoretical expectations. 
Matsumoto {\it et al.}~\cite{Matsumoto:1996zz} found
\begin{equation}
Br( \hbox{o-Ps} \to 5 \gamma)
= \bigl[ 
2.2^{+2.6}_{-1.6} \pm 0.5 \bigr] \times 10^{-6}
\end{equation}
and Vetter and Freedman~\cite{Vetter:2002ps} found
\begin{equation}
Br( \hbox{o-Ps} \to 5 \gamma)
= \bigl[ 
1.67 \pm 0.99 \pm 0.37 \bigr] \times 10^{-6}.
\end{equation}
The energy spectrum for the o-Ps to $3 \gamma$ decay 
is discussed in
\cite{Berestetsky:1982aq,Ore:1949te}
with formulas presented at leading order in $\alpha$.
A first test of the ${\cal O}(\alpha)$ correction 
for the energy spectrum for orthopositronium decay is 
discussed in \cite{Adachi:2015mva}, where the NLO term 
is needed at 92\% C.L.

\subsection{p-Ps decay rate}

For p-Ps the QED prediction is \cite{Kniehl:2000dh,Melnikov:2000fi,Adkins:2003eh} 
\begin{eqnarray}
\Gamma( \hbox{p-Ps} \to 2 \gamma )
&=& 
\frac{\alpha^5 m_e}{2}
\biggl[ 1 + \frac{\alpha}{\pi}
\biggl( \frac{\pi^2}{4} - 5 \biggr)
+ \biggl( \frac{\alpha}{\pi} \biggr)^2
\biggl[
-2 \pi^2 \ln \alpha + 5.1243 (33) \biggr]
\nonumber \\
& & 
\ \ \ + 
\frac{\alpha^3}{\pi} 
\biggl[- \frac{3}{2} \ln^2 \alpha
+
7.9189
 \ln \alpha + \frac{D_p}{\pi^2}
\biggr]
\biggr] + ...
\nonumber \\
\end{eqnarray}
The $4 \gamma$ decay contributes an extra
$
0.274290(8) \bigl( \frac{\alpha}{\pi} \bigr)^2
$
in units of $\frac{\alpha^5 m_e}{2}$ \cite{Adkins:2003eh}.
Summing these contributions gives
\begin{equation}
\Gamma_p = 7989.6178(2) \times 10^6 s^{-1} 
\end{equation}
where we neglect the term proportional to the unknown 
$D_p$ coefficient.
This value compares with the experimental result
\cite{AlRamadhan:1994zz}
\begin{equation}
\Gamma_p = 7990.9 (1.7) \times 10^6 s^{-1}.
\end{equation}
The experimental error 
is 10,000 times the size of the QED theoretical error.

While not needed by present data, 
new physics possibilities might be explored
within the uncertainties 
on the total decay rates, 
presently corresponding to
branching ratios at the ${\cal O} (10^{-4})$ level.
There are strong constraints from other processes
including precision 
measurements of the fine structure constant 
$\alpha$ 
and the electron EDM,
which we discuss in \hbox{Section 3}.

\section{QED tests and $\alpha$ measurements}

The most accurate determinations of $\alpha$ 
come from precision measurements of 
the electron's anomalous magnetic moment,
atom interferometry measurements with Caesium, Cs,
and Rubidium, Rb,
and the Quantum Hall Effect.

The electron anomalous magnetic moment 
$a_e = (g-2)/2$
is generated by radiative corrections,
which have been evaluated to tenth-order
in QED perturbation theory
plus small QCD and weak contributions.
The electron $a_e$ value
gives a precision measurement of $\alpha$ 
(modulo any radiative corrections from 
 new physics beyond the Standard Model).
Atom interferometry experiments with Cs and Rb
and Quantum Hall Effect measurements 
provide a more direct determination
(less sensitive to details of radiative corrections)
but also involve a combination of parameters measured 
in experiments.
The Cs measurements are presently the most accurate.
Comparing these different determinations of $\alpha$ 
gives a precision test of QED
as well as constraining possible new physics scenarios.
Any ``beyond the Standard Model'' effects involving 
new particles active in radiative corrections will 
enter $a_e$ but not the Cs, Rb and Quantum Hall Effect measurements.

The anomalous magnetic moment $a_e$ is related to $\alpha$
through \cite{Aoyama:2017uqe}
\begin{eqnarray}
a_e^{\rm{QED},e} 
&=&
\frac{\alpha}{2 \pi} 
- 0.328 478 965 579 193 ...
\bigl( \frac{\alpha}{\pi} \bigr)^2
+ 
1.181 241 456 587 ...
\bigl( \frac{\alpha}{\pi} \bigr)^3
\nonumber \\
& & 
-1.912 245 764 ...
\bigl( \frac{\alpha}{\pi} \bigr)^4
    +6.675 (192) \bigl( \frac{\alpha}{\pi} \bigr)^5
+ ...
\end{eqnarray}
from Feynman diagrams involving electrons and photons.
Contributions from heavy leptons sum to
\begin{equation}
a_e ({{\rm \hbox{QED: mass-dependent}}})
= 2.747 5719 (13) \times 10^{-12}
\end{equation}
with extra electroweak and QCD corrections
\begin{equation}
a_e^{\rm SM}
=
a_e^{\rm QED} 
+
0.03053 (23) \times 10^{-12} \ ({\rm weak})
\ 
+
\ 
1.6927 (120) \times 10^{-12} \ ({\rm hadronic}).
\end{equation}

The most accurate measurement of $a_e$ comes from the
Harvard group, Gabrielse et al.~\cite{Hanneke:2008tm}
\begin{equation}
a_e^{\rm exp}
=
0.001 159 652 180 73 (28).
\end{equation}
Plans for new, even more accurate, measurements are discussed in
\cite{Gabrielse:2014nha}.
Using the theoretical formulae (10--12),
the $\alpha$ value extracted from $a_e^{\rm SM}$ is
\cite{Aoyama:2017uqe}
\begin{equation}
1 / \alpha|_{a_e^{\rm SM}} = 137.035 999 1491 (331). 
\end{equation}

For direct measurements of $\alpha$, 
the most accurate comes 
from Cs interferometry \cite{Parker:2018sc}
\begin{equation}
1 / \alpha |_{\rm Cs} = 137.035 999 046 (27) 
\end{equation}
with
\begin{equation}
1 / \alpha |_{\rm Rb}
=
137.035 998 996 (85) 
\end{equation}
from $^{87}$Rb atom interferometry
\cite{Bouchendira:2010es,Bouchendira:2013mpa}
quoted by CODATA \cite{Mohr:2015ccw,Mohr:2018pj}.
Note that these Cs and Rb results rely on a number of 
other experimental quantities
involving the Rydberg constant $R_{\infty}$,
the ratio of the atom to electron mass $m_{\rm atom}/m_e$
and new precision measurements of 
the Cs or Rb masses
from recoil of a Cs or Rb atom in an atomic lattice,
{\it viz.}
\begin{equation}
\alpha^2
=
\frac{2 R_{\infty}}{c}
\frac{m_{\rm atom}}{m_e} \frac{h}{m_{\rm atom}}.
\end{equation}
Here $c$ is the speed of light and $h$ is Planck's constant.
Quantum Hall Effect 
experiments yield the value (CODATA \cite{Mohr:2015ccw})
\begin{equation}
1 / \alpha |_{\rm QHE}
=
137.036 0037 (33).
\end{equation}
The Hall conductivity of two dimensional electron 
systems is quantised in integral multiples of 
$e^2/h$ \cite{klitzing}.
Interest in the ratio $e^2/h$ lies not only in its 
application as an ``atomic'' resistance standard based 
on fundamental constants but also as a method to measure $\alpha$.
The different measurements of $1/\alpha$ are collected in Table 1.

\begin{table}[t!]
\begin{center}
\begin{tabular}[t]{c|ll}
\hline
Process & $1 / \alpha$  & Reference \\
\hline
$a_e^{\rm SM}$ &
137.035 999 1491 (331)  & \cite{Aoyama:2017uqe}
\\
$h/m (^{133} {\rm Cs})$ & 
137.035 999 046 (27)    & \cite{Parker:2018sc}
\\
$h/m (^{87} {\rm Rb})$
& 137.035 998 995 (85)  & \cite{Mohr:2018pj}
\\
QHE & 137.036 003 7(33) & \cite{Mohr:2015ccw}
\\
\hline
\end{tabular}
\end{center}
\caption{
Values of $\alpha$ extracted from different experiments.
}
\label{bagparam}
\end{table}

The new most accurate Cs atomic physics measurement
corresponds to 
\begin{equation}
a_e^{\rm exp} - a_e^{\rm th}|_{\rm Cs}
=
(-88 \pm 36 ) \times 10^{-14}
\end{equation}
when we substitute the $\alpha$ value in Eq.~(15)
into Eqs.~(10--12) 
to obtain the value $a_e^{\rm th}|_{\rm Cs}$.
Suppose we interpret this 2.5 $\sigma$ ``discrepancy'' 
as an upper limit on contributions coming from new physics,
$\Delta a_e^{\rm New \ Physics} 
\equiv a_e^{\rm exp} - a_e^{\rm th}$.
If this originates from new heavy exchanges with
coupling constant $g_X$ and mass scale $\Lambda$,
then
\begin{equation}
| \Delta a_e^{\rm New \ Physics} | 
\approx 
\frac{1}{2\pi} \frac{g_X^2}{4 \pi} m_e^2 \ / \ \Lambda^2 
\end{equation}
fixes $\Lambda$ bigger than $\sim 40$ GeV
(much below collider constraints)
with
$\frac{g_X^2}{4 \pi} \sim 4/137 \simeq 0.03$ 
(that is, 
 taking coupling constants of order the Standard Model ones).
If one instead assumes a new light particle with mass 
$m_X^2 \ll m_e^2$, then \cite{Gninenko:2002jn}
\begin{equation}
| \Delta a_e^{\rm New \ Physics} | 
\approx 
\frac{1}{2\pi} \frac{g_X^2}{4 \pi} .
\end{equation}
Taking the numbers in Eq.~(19), the upper bound on 
$g_X^2/4\pi$ is about 
6 $\times 10^{-12}$, 
a factor of 50 reduction 
from the earlier analysis in ~\cite{Gninenko:2002jn}.
This upper bound constrains
the branching ratios
for possible decays of o-Ps
to a photon and very-light mass pseudoscalar 
and to two photons and a new light vector boson 
to be 
$< 10^{-6}$ and $< 10^{-9}$ respectively. 
Further constraints
on pseudoscalar axion models
come from astrophysics and laboratory experiments,
see Section 4.3 below.
In a recent paper \cite{Davoudiasl:2018fbb} 
the difference in Eq.(19) and the muon $g-2$ anomaly
\cite{Bennett:2006fi}
are interpreted together
through introduction of a new scalar with mass 
bigger than about 250 MeV
and couplings 
to the muon and electron 
of $\sim 10^{-3}$ and a few times $10^{-4}$.

New, most accurate, measurements of the electron EDM \cite{Andreev:2018ayy}
give
\begin{equation}
| d_e | < 1.1 \times 10^{-29} e {\rm cm}.
\end{equation}
A finite value of $d_e$ 
would correspond to some new $CP$ violating interaction.
Within typical extensions of the Standard Model
involving new heavy particles,
the electron EDM constraint puts limits on the mass scales
of this new physics as 
30 TeV in one-loop calculations and 3 TeV at two-loops,
strong constraints on new physics models
which are competitive with the constraints from the LHC
\cite{Andreev:2018ayy}.
(These numbers are obtained assuming similar size couplings 
 to the Standard Model ones and 
 $\sin \phi_{\rm CP} \sim 1$,
 where $\phi_{\rm CP}$ is the $CP$ violating phase).
Changing from possible new heavy particles
to exchanges involving new near-massless particles
corresponds to an upper bound on their coupling to
electrons of
$g_X^2/4\pi \sim 8 \times 10^{-18}$ 
in the leading-order calculation and
$\sim 5 \times 10^{-9}$ within two-loop calculations
for $CP$ violating interactions
(evaluated by rescaling the heavy mass scale in the
 calcuations to the electron mass and
 assuming no special phase cancellation in the EDM).
The latter bound on $g_X^2/4\pi$ 
corresponds to an upper bound on the 
branching ratio 
for $CP$ violating o-Ps decays of about $10^{-9}$.

\section{Rare and exotic decays and new physics}

\subsection{$C$ and $P$ violating decays}

Experimental bounds on possible 
$C$-violating decays of positronium
have been reported by earlier experiments
\begin{eqnarray}
& & 
BR ( \hbox{p-Ps} \to 3 \gamma / \hbox{p-Ps} \to 2 \gamma )
< 2.8 \times 10^{-6} \ {\rm at} \ 68\% {\rm C.L.} 
\ \ \cite{X5}
\\
& &
BR ( \hbox{o-Ps} \to 4 \gamma / \hbox{o-Ps} \to 3 \gamma )
< 2.6 \times 10^{-6} \ {\rm at} \ 90\% {\rm C.L.} 
\ \ \cite{X4}
\\
& & 
BR ( \hbox{p-Ps} \to 5 \gamma / \hbox{p-Ps} \to 2 \gamma )
< 2.7 \times 10^{-7} \ {\rm at} \ 90\% {\rm C.L.} 
\ \ \cite{X2}
\end{eqnarray}
J-PET will push these limits.
With a 10 MBq positronium source and 
upgraded 4 layer detector geometry, one expects 
to measure
$9.4 \times 10^{10}$ o-Ps to 3 photon decays and 
$3 \times 10^{11}$ p-Ps to two photon decays in 
365 days of data taking 
\cite{pmpriv}.

QED forbidden decays can proceed through weak interactions 
but with very small branching ratios 
because of the massive W and Z boson propagators that 
appear in these reactions.
For example, 
the three photon decay of p-Ps is forbidden in QED 
but can occur through weak interactions involving 
a W-boson loop with branching ratio
$
BR (\hbox{p-Ps} \to 3 \gamma) = 4.4 \times 10^{-77}
$
\cite{Pokraka:2017ore}.
Tiny branching ratios are found for decays into a 
photon and two neutrinos~\cite{Pokraka:2016jgy},
less than about $10^{-21}$ for o-Ps and $10^{-24}$ for p-Ps. 
Exotic decays to a single photon and possible light mass 
``dark photon'' (dark matter candidate)
have been postulated with branching ratio 
up to ${\cal O}(10^{-10})$ \cite{Perez-Rios:2018veb}.

\subsection{CP violation}

After the electron EDM, 
Eq.~(22),
o-Ps decays are an experimentally clean 
system to look for $CP$ violation with charged leptons 
\cite{Bigi:2000yz}. 
The observed matter antimatter asymmetry in the Universe
requires some extra source of $CP$ violation beyond the
quark mixing described by the Cabbibo-Kobayashi-Maskawa
(CKM) matrix in the electroweak Standard Model.
Recent measurements by the T2K Collaboration in Japan
are consistent with $CP$ violation 
in the neutrino sector at
the level of two standard deviations \cite{Abe:2018wpn}.
So far there is no hint from experiments for $CP$ violation
with charged leptons.
The electron EDM places strong constraints on any new effect.
Searching for $CP$ violation in 
positronium decays is an active topic of investigation.
To see a signal new effects will need to be much larger
than Standard Model $CP$ violations, which are very much
suppressed.
%

Spin-one o-Ps decays are sensitive to $CP$ and $CPT$ odd correlations through the spin vector of the o-Ps.
Previous experiments have focused on the correlations
\begin{eqnarray}
A_{CP} &=&
\langle ( \vec{S} . \vec{k}_1 ) 
        ( \vec{S} . ( \vec{k}_1 \times \vec{k}_2 ) 
\rangle 
\\
A_{CPT} &=& 
\langle \vec{S} . (\vec{k}_1 \times \vec{k}_2) \rangle
\end{eqnarray}
measuring the $T$-odd integrated moments between the
polarisation vector $\vec{S}$ of the o-Ps and the momenta of the emitted photons with magnitude $k_1 \geq k_2 \geq k_3$.

New $CP$ and $CPT$ observables enter 
if we can also measure observables related to
Compton scattering of photons from the positronium decay
in the detector~\cite{Moskal:2016moj}.
The Compton scattering cross section is peaked 
perpendicular to the electric field and 
polarisation axis of 
the incident photon \cite{Moskal:2018pus,Klein:1929}.
This leads to defining the polarisation related quantities
$\vec{\epsilon_i} = \vec{k_i} \times \vec{k_i}' / k_i k_i'$, 
where $\vec{k_i}$ and $\vec{k_i}'$ 
are the momenta of a photon from the positronium 
decay and the rescattered photon 
from Compton scattering in the detector~\cite{Moskal:2016moj}.
These $\vec{\epsilon_i}$ vectors are peaked 
to lie along the axis of the 
incident photon polarisation vector.
They are even under $P$ and $T$ transformations.
One can form new $CP$ and $CPT$ correlations between 
the o-Ps spin vector, the momenta of the radiated 
photons and the $\vec{\epsilon_i}$ vectors, see Table 2.
The first three rows in Table 2
refer to correlations involving
just the o-Ps decay and the second three rows 
involve correlations between the positronium system 
and Compton scattering processes in the detector.

We briefly comment on the relation between the
$\vec{\epsilon_i}$ and the polarisation vectors 
$\varepsilon_{\mu}^{(j)}$ 
for circularly polarised photons \cite{bjd,Sozzi}.
These polarisation vectors transform under $P$ as
$\vec{\varepsilon}^{\ (3)} = {\vec k}/|k| 
 \to - \vec{\varepsilon}^{\ (3)}$ 
for longitudinally polarised photons
and
$\vec{\varepsilon}^{\ L}(\vec{k}) = \vec{\varepsilon}^{\ R}(-\vec{k})$
for left- and right-handed 
circularly polarised photons with definite helicity.
That is, parity transformations 
change the sense of rotation about the flipped momentum axis. 
Of the two transverse direction components, 
one is parity even, like $\vec{\epsilon_i}$, 
with the other parity odd.

The present most accurate measurements are
\begin{eqnarray}
C_{CP} &=&  0.0013 \pm 0.0022, \ \ \ \ \ 
{\rm Ref.~\cite{Yamazaki:2009hp}}
\nonumber \\
C_{CPT} &=& 0.0071 \pm 0.0062, \ \ \ \ \ 
{\rm Ref.~\cite{X1}}.
\end{eqnarray}
J-PET aims to improve the accuracy here to ${\cal O}(10^{-5})$ 
for $CP$ and $CPT$ 
observables~\cite{Moskal:2016moj,Eryk:2017}.
Experiments to improve the accuracy on CP observables
are also planned at MIT~\cite{mitps}.

Standard Model QED final state interactions can mimic
$CP$, $T$ and $CPT$ violation by inducing finite 
values of the correlations in Eqs.~(26, 27) at the 
level of ${\cal O}(10^{-9}) - {\cal O}(10^{-10})$ 
\cite{Bernreuther:1988tt,Y2}.
$CPT$ symmetry is a fundamental property of 
relativistic quantum field theories like QED~\cite{bjd}.
While QED interactions preserve $CPT$, o-Ps as an
unstable state is not an eigenstate of 
time reversal symmetry $T$ and of $CPT$.
(It is an eigenstate of $C$ and $CP$.)
This leads to non-vanishing values of the correlations 
in Eqs.(26) and (27) in detailed calculations of the 
final state interactions with the leading contribution 
coming from 
light by light scattering of two of 
the three photons in 
the final state~\cite{Bernreuther:1988tt}.

\begin{table}[t!]
\begin{center}
\begin{tabular}[t]{c|lllll}
\hline
Operator & $C$ & $P$ & $T$ & $CP$ & $CPT$ \\
\hline
$\vec{S} \cdot \vec{k}_1$ & + & - & + & - & - \\
\hline
$\vec{S} \cdot (\vec{k}_1 \times \vec{k}_2)$ & + & + & - & + & - \\
\hline 
$(\vec{S} \cdot \vec{k}_1) (\vec{S} \cdot (\vec{k}_1 \times \vec{k}_2))$ & + & - & - & - & + \\ 
\hline
$\vec{k}_1 .
\vec{\epsilon}_2$ & + & - & - & - & + \\
\hline
$\vec{S} \cdot \vec{\epsilon}_1$ & + & + & - & + & - \\
\hline
$\vec{S} \cdot (\vec{k}_2 \times \vec{\epsilon}_1)$ & + & - & + & - & - \\
\hline
\end{tabular}
\end{center}
\caption{
Operators for the $\hbox{o-Ps} \to 3\gamma$ process, and their properties with respect to $C$, $P$, $T$, $CP$ and $CPT$ symmetries. 
Here $\vec{k_1}$ and $\vec{k_2}$ denote momentum 
vectors of photons ordered according to their magnitude,
$k_1 \geq k_2$,
$\vec{S}$ is the spin vector for the o-Ps, 
and
$\vec{\epsilon_i} = \vec{k_i} \times \vec{k_i}' / k_i k_i'$ 
where $\vec{k_i}$ and $\vec{k_i}'$ denote the 
momentum of the $i^{\rm th}$ photon before and after 
Compton scattering in the detector~\cite{Moskal:2016moj}.
}
\end{table}

The mass scale for weak interference effects in positronium 
is given by 
$G_F m_e^2 \sim 10^{-11}$,
so one needs a dramatic enhancement to obtain an observable effect.
Theoretical conditions needed for observation of $CP$
violation in positronium decays are discussed in 
\cite{Bernreuther:1988tt}.
The electron EDM measurements strongly constrain any 
new sources of $CP$ violation coupled to the electron. 
Any observation of $CP$ violation in positronium decays
in the next generation of experiments
would point to some cancellation of $CP$ phases in the EDM.

\subsection{Visible exotic decays of o-Ps to a photon and
axion}

A possible resolution of the strong $CP$ puzzle in QCD 
is to postulate the existence of a new very-light mass pseudoscalar called the axion 
\cite{Weinberg:1977ma}
which couples through the Lagrangian term
\begin{equation}
{\cal L}_{a} =
- \frac{1}{2} \partial_{\mu} a \partial^{\mu} a
+ 
\frac{a}{M} 
\frac{\alpha_s}{8 \pi} G_{\mu \nu} {\tilde G}^{\mu \nu}
+ 
\frac{a}{M} 
\frac{\alpha}{3 \pi} F_{\mu \nu} {\tilde F}^{\mu \nu}
+ 
\frac{i f_\psi}{M} \partial_{\mu} a
\ \bar{\psi} \gamma^{\mu} \gamma_5 \psi - ...
\end{equation}
Here the second and third terms denote 
coupling to gluons and photons, and
the term in $\psi$ denotes possible fermion couplings 
to the axion $a$ with $f_\psi \sim {\cal O}(1)$.
The mass scale $M$ plays the role of the axion decay constant 
and sets the scale for this new physics.

One finds \cite{Bernreuther:1981ah,Cleymans:1983ci} 
\begin{equation}
\Gamma ( \hbox{o-Ps} \to \gamma a )
=
\frac{8}{3} \ g_a^2 \ \alpha \ | \Psi_n (0) |^2 \
\frac{m_{\hbox{o-Ps}}^2 - m_a^2}{m_{\hbox{o-Ps}}^4} 
\end{equation}
with $g_a = - 2 m_e f_e / M$. 
Here $\Psi_n (0)$ is the wave function at the origin,
\begin{equation}
| \Psi_n (0) |^2 = ( \alpha m_e )^3 /  8 \pi n^3
\end{equation}
where $n = 1$ for the ground state
\cite{Mikaelian:1978xm};
$m_e$, $m_a$ and $m_{\hbox{o-Ps}}$ 
are the electron, axion and positronium masses.
Eq.~(30) becomes
\begin{equation}
\Gamma ( \hbox{o-Ps} \to \gamma a )
=
\frac{1}{3 \pi} \ \alpha^4 \ m_e^3 \frac{f_e^2}{M^2}
\biggl( 1 - \frac{m_a^2}{m_{\hbox{o-Ps}}^2} \biggr).
\end{equation}
Axions are possible dark matter candidates.
Constraints from experiments tells us that $M$ must 
be very large.
Laboratory based experiments 
together with astrophysics and cosmology constraints 
suggest a favoured QCD axion mass between 
$1 \ \mu$eV and 3 meV 
\cite{Kawasaki:2013ae,Baudis:2018bvr}
corresponding to 
$M$ between $6 \times 10^9$ and $6 \times 10^{12}$ GeV.
Taking $f_e \sim {\cal O}(1)$ and dividing 
by the leading order 
decay rate term from QED in Eq.~(1)
gives an expected branching ratio for the decay to a 
photon and axion between about $10^{-28}$ and $10^{-22}$.

In order to see a signal with branching ratio of
${\cal O}(10^{-8})$ one would need $M$ close to the TeV scale.
The axion mass $m_a \sim 1/M$.
If the axion is too heavy it would lead to axions 
carrying  too much energy out of supernova explosions,  
thereby observably shortening the neutrino arrival 
pulse length recorded on Earth in contradiction to
Sn 1987a data \cite{Kawasaki:2013ae}.

\subsection{Invisible decays}

The search for invisible decays is an interesting topic of experimental investigation.
Invisible decays can occur in mirror matter models \cite{Glashow:1985ud}.
Here, the o-Ps can oscillate into a virtual photon
which then oscillates 
into an invisible ``mirror photon'' and
``mirror positronium'' 
(with no interaction with the detector).
Mirror matter was first proposed in connection with 
parity violation.
The idea is that under spatial inversion particles 
should transform into parity reflected 
new mirror states \cite{Lee:1956qn},
which then restore parity symmetry in nature.
The mirror particles would interact with normal particles mainly through gravity and, 
as such, are dark matter candidates.
Oscillations between photons and their mirror partners 
could proceed through the interaction term 
\cite{Glashow:1985ud}
\begin{equation}
{\cal L} = 
\epsilon \ F^{\mu \nu} F_{\mu \nu}^m.
\end{equation}
The upper limit on the mixing term
deduced from successful prediction of
the primordial $^4$He abundance by the Standard Model 
is \cite{Carlson:1987si}
\begin{equation}
\epsilon \leq 3 \times 10^{-8}.
\end{equation}
A value of $\epsilon$ between
$10^{-10}$ and $4 \times 10^{-9}$ 
has been suggested in 
models aiming to explain the DAMA anomaly 
in dark matter physics
\cite{Cerulli:2017jzz,Foot:2014mia}.

The most accurate constraint on a possible invisible decay 
of o-Ps in vacuum comes from the 
ETH Z\"urich group~\cite{Vigo:2018xzc} 
\begin{equation}
BR ( \hbox{o-Ps} \to {\rm invisible}) < 5.9 \times 10^{-4},
 \ \ \ \  90\% \ {\rm C.L.}
\end{equation} 
which is interpreted as a constraint on 
the mixing parameter, $\epsilon \lesssim {\cal O}(10^{-7})$.
For experiments in medium 
one also has to take into account o-Ps collisions 
within the apparatus which act as a source of decoherence 
and dilute the accuracy of 
the measurement of $\epsilon$ \cite{Badertscher:2006fm}.
Next generation measurements aim to probe possible 
branching ratios of ${\cal O}(10^{-8})$ corresponding 
to mixing with
$
\epsilon \sim 4 \times 10^{-9}$~\cite{Crivelli:2010bk}.
These experiments will use a higher positron flux 
with improved confinement cavity and tagging system.

\section{Conclusions}

Precision measurements of positronium decays test 
fundamental symmetries with charged leptons.
The large relative error on the experimental decay rates 
in Eqs.~(3,~4) for o-Ps and Eq.~(9) for p-Ps compared 
to the QED theory uncertainties in Eqs.~(2,~8) demands 
improved experimental accuracy.
Constraints from other observables involving the QED 
fine structure constant tell us that 
any new physics effects will be small.
The branching ratios for processes beyond the most 
simple QED decays of o-Ps and p-Ps
most likely start at ${\cal O}(10^{-6})$ 
with the five photon decay of o-Ps and 
the four photon decay of p-Ps. 
QED final state effects can mimic $CP$, $T$ and $CPT$ 
violation 
in spin momentum correlations measured in o-Ps decays
at the level of 
${\cal O}(10^{-9}) - {\cal O}(10^{-10})$.
This effect is associated with the fact that 
o-Ps as an unstable state is not an eigenstate of $T$. 
The range is similar to the upper bound ${\cal O}(10^{-9})$ 
suggested by measurements of the electron EDM 
for the branching ratio for o-Ps decays  
involving possible new $CP$ violating interactions
(with the EDM measurement interpreted 
 assuming no special phase cancellation in the EDM).
The branching ratio for o-Ps decays 
to an axion and single photon is expected 
to be below ${\cal O}(10^{-22})$.
For testing models of possible mirror particles 
the aim is to be sensitive to branching ratios 
for possible invisible decays at or below ${\cal O}(10^{-8})$.

\section*{\bf Acknowledgements}

I thank
C. Aidala, P. Crivelli, N. Krawczyk, W. Krzemien, 
P. Moskal, J. Raj and M. Silarski for stimulating discussions.
This article is based on a lecture presented at the 
3rd Symposium on Positron Emission Tomography, Cracow, September 10-13 2018.

\end{document}